\documentstyle[aps,twocolumn]{revtex}
\begin{document}
\voffset 0.8truecm
\title{A note on separability criteria for multipartite state
}
\author{
Heng Fan
}
\address{
Quantum computing and information project,
ERATO, 
Japan Science and Technology Corporation,\\
Daini Hongo White Building.201, Hongo 5-28-3, Bunkyo-ku, Tokyo 133-0033, Japan.\\
}
\maketitle
                                                        
\begin{abstract}
The recent proposed realignment separability criterion for
mixed is analyzed. We identify the essential part of this criterion
is a swap operator followed by a partial transposition.
Then we analyze the separability criterion of permutation indices
of the density matrix for multipartite state which is a generalization
of the realignment criterion. We give a method to solve the
equivalence problem of the separability criterion of permutation indices
of the density matrix. For tripartite state, we show the non-trivial
separability criteria are either partial transposition criterion
or realignment criterion.
\end{abstract}
       
Since the partial transposition criterion for separability of mixed
states proposed by Peres \cite{Peres}
and proved to be necessary and sufficient
condition for $2\times 2$ and $2\times 3$ systems
by Horodeckis (M,P,R)\cite{HHH96},
no other operational separability criterion appeared for a long time.
Recently, a new criterion for separability (realignment criterion)
was found by Rudolph\cite{Rudolph} and Chen and Wu\cite{CW1}.
This criterion was shown to be strong
enough to detect genuinely tripartite entanglement,
and generalized to multipartite system\cite{HHH02}. And a generalized
partial transposition was also proposed to 
multipartite system\cite{CW2}.

In this short note, we point out that the essential part of
realignment criterion for separability is actually
a one side swap operator plus a partial tranposition.
Then it can be shown that 
for bipartite and tripartite systems, the
non-trivial criteria for separability are only
partial transposition criterion
or realignment criterion.
But for 4 or more-partite system, the partial transposition criterion
and the realignment criterion combined together can give new
separability criteria. For multipartite system, it was pointed
out that any permutation of indices of density matrix leads
to separability criterion\cite{HHH02}. Though many of them
are non-equivalent to each other, but the problem of equivalence
of those criteria is still open. In this short note,
we point out that we do not need to perform all
permutation of indices of density matrix to check whether
the system is entangled or not,
the one side swap operators followed by partial transpositions
will provide all non-trivial criteria for separability.
Instead of the swap operator,
any non-local operators which can keep the tensor product form
followed by a partial transposition will provide a separability
criterion.

The original realignment separability criterion \cite{Rudolph,CW1}
can be described
as follows: For a separable bipartitie density matrix $\rho _{AB}$,
the trace norm of the matrix ${\cal {R}}(\rho _{AB})$
which is obtained by
performing the realignment of entries of the density
matrix $\rho _{AB}$ should satisfy
$||{\cal {R}}(\rho _{AB})||\le 1$, where
the trace norm is defined as the sum of singular values of the
matrix ${\cal {R}}(\rho _{AB})$.
So, if
$||{\cal {R}}(\rho _{AB})|| >1$, $\rho _{AB}$ is entangled.
The realignment of a density matrix act on the basic operators
of the from $|i_1\rangle \langle i_1'|\otimes
|i_2\rangle \langle i_2'|$ leads to
$|{i_1'}^*\rangle \langle i_2'|\otimes
|i_1\rangle \langle i_2^*|$.
This realignment can be realized by the swap operators and
partial transposition as follows:
\begin{eqnarray}
|{i_1'}^*\rangle \langle i_2'|\otimes
|i_1\rangle \langle i_2^*|=
V_{12}^L\left( (|i_1\rangle \langle i_1'|\otimes
|i_2\rangle \langle i_2'|)V_{12}^R\right) ^{t_2}.
\label{realig}
\end{eqnarray}
where superindices $t_2$ means partial transposition on
the second space, 
$V_{12}^{L,R}$ are left (right) hand side of
sway operators defined as
$V_{12}=\sum _{jkj'k'}\delta _{jk'}\delta _{j'k}|j\rangle \langle j'|\otimes
|k\rangle \langle k'|$,
$j,k'=1, \cdots, d_2, j',k=1,\cdots, d_1$, it is the right hand side
swap operator for $C_{d_1}\otimes C_{d_2}$, and left hand side
swap operator for $C_{d_2}\otimes C_{d_1}$ .
We have $V_{12}^{L}=(V_{12}^{R})^t=(V_{12}^{R})^{-1}$,
where $t$ means a whole transposition on both the first and second
spaces, and we know the swap operators are
unitary operators. For Hilbert spaces $C_{d_1}\otimes C_{d_2}$,
if $d_1=d_2$, we have $V_{12}^{L}=V_{12}^{R}$. 

We know the singular values of the matrix
${\cal {R}}(\rho _{AB})$
is invariant if we do not act the right hand side swap
operator presented in (\ref{realig}).
So, we can find the essential part of the realignment
separability criterion proposed in Ref.\cite{Rudolph,CW1}
is actually a single side swap operator followed by
a partial transposition as presented as follows,
\begin{eqnarray}
\left( (|i_1\rangle \langle i_1'|\otimes
|i_2\rangle \langle i_2'|)V_{12}^R\right) ^{t_2}.
\label{realig1}
\end{eqnarray}
In what follows, we also call a single side swap
operator followed by a partial transposition as
realignment criterion.
It is pointed out in Ref.\cite{HHH02,Zuk}
that any permutation of indices of density matrix
lead to separability criterion.
It is clear that any permutation of indices of density
matrix can be realized by swap operators and
partial transpositions. In case that there are a lot
of indices permutations, and a lot of them are
equivalent though we expect many of them are
non-equivalent. We can check explicitly which
of them are equivalent or non-equivalent by
looking how they can be realized by swap operators
and partial transpositions. 

For a bipartite system,
the permutation of indices of density matrix lead
only two non-trivial separability criterion:
partial transposition criterion and realignment
criterion.

Let's now analyze the tripartite system.
It is known that the realignment criterion can detect
genuinely tripartite entanglement.
Here we would like to check besides partial
transposition criterion and realignment criterion, whether
the permutation of indices of density matrix can provide
some new separability criteria.
The basic operators of the tripartite density matrix is
as follows:
\begin{eqnarray}
|i_1\rangle \langle i_1'|\otimes
|i_2\rangle \langle i_2'|\otimes
|i_3\rangle \langle i_3'|.
\end{eqnarray}
The partial transposition method 
$t_1, t_2, t_3$ will provide 3 non-equivalent separability
criteria. The transposition in 2 Hilbert spaces is
equivalent to partial transposition in 1 Hilbert space
because the singular values of a matrix are invariant
under a whole transposition. Now, let's see the
realignment criteria, we have the following
non-equivalent realignment:
\begin{eqnarray}
\left( (|i_1\rangle \langle i_1'|\otimes
|i_2\rangle \langle i_2'|\otimes
|i_3\rangle \langle i_3'|)V_{jk}^{R}\right) ^{t_l},
\nonumber \\
l=j~{\rm or}~k, ~j\not= k=1,2,3.
\label{tripar}
\end{eqnarray}
Here we remark, for bipartite system, after the swap
operator $V_{12}^R$, partial transposition $t_1$ and
$t_2$ are equivalent in the sense the singular
values are the same because $t_1$ can be deduced from
$t_2$ by a whole transpose $t=t_1t_2$.
For tripartite system, $V_{jk}^R$
followed by $t_j$ or $t_k$ give different results
since they cann't be deduced from each other by
a whole transpose.
So, we have 6 non-equivalent realignment criteria
for tripartite system.
Now, let's see whether we have anything more.
Apparently, swap operators themselves can not
provide separability criteria.
Only the partial transposition can lead to
some separability criteria.
And the swap operator is unitary, so, we can apply
the partial transposition in the final step.
The fact that partial transposition in two
Hilbert spaces is equivalent partial transposition
in only one Hilbert space for tripartite state
leads to that we only need to consider
the partial transposition in one Hilbert space.
After these arguments, we know the only possible new
separability criteria left is several swap operators
followed by partial transposition in one Hilbert space
other than (\ref{tripar}).
Next, we show that the one side, for example, right hand
side, swap operators followed by partial transposition
can provide all non-equivalent separability criteria.
The argument for tripartite system is also applicable
for other multipartite system. For any permutation
indices of density matrix, we can move by swap operators,
which do not change the singular values,
the state $|i_1\rangle ,|i_2\rangle ,|i_3\rangle $ to
their original Hilbert spaces $C_{d_j}$
either as the form $|i_j\rangle \langle \cdot |$ or
$|\cdot \rangle \langle i_j^* |$. Thus, only
right hand side swap operators followed by partial
transposition are enough to obtain the form
$|i_j\rangle \langle \cdot |$ or
$|\cdot \rangle \langle i_j^* |$ in $C_{d_j}$.
Finally, we show several right hand side swap
operators followed by partial transposition do not
lead to new criteria other than (\ref{tripar}).
We have the following cases
\begin{eqnarray}
(|i_1\rangle \langle i_2'|\otimes
|i_2\rangle \langle i_3'|\otimes
|i_3\rangle \langle i_1'|)^{t_j},
\nonumber \\
(|i_1\rangle \langle i_3'|\otimes
|i_2\rangle \langle i_1'|\otimes
|i_3\rangle \langle i_2'|)^{t_j},
\nonumber \\
~j=1,2,3.
\label{tripar1}
\end{eqnarray}
It can be checked easily, all these cases are equivalent
to the cases (\ref{tripar}), for example,
$$(|i_1\rangle \langle i_2'|\otimes
|i_2\rangle \langle i_3'|\otimes
|i_3\rangle \langle i_1'|)^{t_1}$$
is equivalent to
the case
$$\left( (|i_1\rangle \langle i_1'|\otimes
|i_2\rangle \langle i_2'|\otimes
|i_3\rangle \langle i_3'|)V_{12}^{R}\right) ^{t_1}.$$
For tripartite state, permutation indices of density
matrix will provide $6!-1=719$ possible
separability criteria. Here we show only 9 of them
are non-equivalent which either are the partial
transposition criteria or realignment criteria.

For multipartite system, permutation of indices of
density matrix will provide some new separability
criteria other than partial transposition criteria
or realignment criteria. A simple example for
four-partite system is as follows:
\begin{eqnarray}
\left( (|i_1\rangle \langle i_1'|\otimes
|i_2\rangle \langle i_2'|
\otimes |i_3\rangle \langle i_3'|\otimes
|i_4\rangle \langle i_4'|
)V_{12}^R\right) ^{t_2t_3}.
\label{4partite}
\end{eqnarray}
The permutation of indices of the density
matrix is actually a realignment in $C_{d_1}\otimes C_{d_2}$
plus a partial transposition in $C_{d_3}$.
This permuation
of indices of density matrix can not be realized by solely
partial transposition or realignment.
A whole
transpostion of relation (\ref{4partite}) will leads
to another realignment in $C_{d_1}\otimes C_{d_2}$
plus a partial transpostion in $C_{d_4}$.

Since permutation indices of density matrix gives
a lot of separability criteria. It is necessary
to find the non-equivalent criteria as we do for
tripartite system. For multiparite system, it is
also true that one side, for example, right hand side swap operators
followed by partial transposition will give
all non-equivalent separability criteria as we argued for
tripartite system.
For N-partite system, the partial transposition can
be limited to $[N/2]$ subspaces, where $[N/2]$ is the
integer part of $N/2$.

Next, we present a method to check whether one permutation
indices of density matrix is equivalent to another.
Since applying swap
operators, left or right hand sides, on the matrix
obtained by permuation indices
does not change
the singular values of a matrix, so all forms which can
be realized by swap operators are equivalent.
We give the following method to check equivalent problem:
1, If the parity change is caused by partial transposition
in more than $[N/2]$ subspaces, apply a whole transpositon.
2, By swap operators, move all states without
parity change to first two parts,
and try to keep the original form, for exampel, if we have
$|i_j\rangle $ and $\langle i_j'|$, let them in the form
$|i_j\rangle \langle i_j'|$ and list them in first part,
otherwise in alphabet order in the second part like
the following
\begin{eqnarray}
&&|i_{j_1}\rangle \langle i'_{j_1}|\otimes
\cdots
\otimes |i_{j_m}\rangle \langle i'_{j_m}|
\nonumber \\
&&~~\otimes |i_{j_{m+1}}\rangle \langle i'_{j'_{m+1}}|
\otimes \cdots     
\otimes |i_{j_{n}}\rangle \langle i'_{j'_{n}}|,
\end{eqnarray}
where $j_{m+1}<j_{m+2}<\cdots <j_n$,
$j'_{m+1}<j'_{m+2}<\cdots <j'_{n}$. We know now the
first part indices does not have any change, the
second part are caused by realignment.
3, Similary, arrange the states with parity change to third
and fourth part, the third part are caused by solely partial
transposition while the fourth part are realigment.
\begin{eqnarray}
&&|{i'_{j_{n+1}}}^*\rangle \langle i_{j_{n+1}}^*|\otimes
\cdots
\otimes |{i'_{j_k}}^*\rangle \langle i_{j_k}^*|
\nonumber \\
&&~~\otimes |{i'_{j'_{k+1}}}^*\rangle \langle i_{j_{k+1}}^*|
\otimes \cdots 
\otimes |{i'_{j_{N}}}^*\rangle \langle i_{j'_{N}}^*|.
\end{eqnarray}
After these 3 steps, we can find easily whether
two permutation indices are equivalent or not.

One can find the swap operator keep the tensor product form.
Instead of the swap operator, if there are any other non-local
operators which keep the tensor product form, we essentially
can find a new separability criteria by applying these
operators solely or followed by partial transposition.
However, the swap operator is the only non-local unitary
operator which keep the tensor product form\cite{KC,keiji}.
So, the only chance is to find a non-unitary operator which
keep the tensor product form. This kind of operator
solely can provide a possible separability criterion,
and apply the non-unitary operator followed by partial
transposition will be another separability criterion.
Since partial transposition itself is a non-unitary operator
which keeps the tensor product form, we can see that
any two, or several operators which keep the tensor product
form combined together can provide a separability criterion.
In short, any non-unitary operator which keeps the
tensor product form is a separability criterion by
checking whether the trace norm of the new matrix
is larger than 1.

Chen and Wu also generalized their results of realignment
criteria from bipartite system to multipartite system\cite{CW2}.
For multipartite system, this generalization is at least
as powerful as the permutation indices of the density
matrix proposed by Horodeckis (M.P.R) \cite{HHH02}.
However, it not clear whether this method is
more powerful than the permutation indices method.
In this paper we study the equivalence problem of these criteria.

In summary, we analyze the equivalence problem of
the recent proposed separability criterion by using
the permutation indices of the density matrices.
Since permutation indices will provide a lot of
separability criteria, it is necessary to identify
which of them are equivalent or non-equivalent.
Mainly we us the fact that the realignment criterion
and the permutation indices method can be
realized by two basic operations: swap opterator
and partial transposition.
And also we point out that any non-unitary operator
which keep the tensor product form will provide
a separability criterion.
 
{\it Acknowlegements}: The author would like to
thank P.Horodecki and K.Matsumoto for very useful
discussions. The author also would like to
thank K.Chen and L.A.Wu for explaining
their work to the author and for communications.

\end{document}